\documentclass[aip, jmp, showpacs, showkeys]{revtex4-1}
\usepackage{amsfonts}
\usepackage{amssymb}
\usepackage{amstext}
\usepackage{amsmath}
\usepackage{latexsym}
\usepackage{color}

\newtheorem{defi}{Definition}[section]
\newtheorem{theorem}[defi]{Theorem}
\newtheorem{remark}[defi]{Remark}

\newcommand{\betheo}{\begin{theorem}$\!\!${\bf \,\,\,}}
\newcommand{\entheo}{\end{theorem}}

\DeclareMathOperator{\okr}{{\stackrel{{\scriptscriptstyle{\mathsf{def}}}}{=}}}

  \def\ulamek#1#2{\mbox{\normalfont$\frac{#1}{#2}$}}

\catcode `\@=11 \@addtoreset{equation}{section}

\catcode `\@=12

\begin{document}

\title[Squeezed States and Hermite polynomials in a Complex Variable]
{Squeezed States and Hermite Polynomials in a Complex Variable}

\author{S. T. Ali}
\email{twareque.ali@concordia.ca}

\affiliation{Department of Mathematics and Statistics, Concordia University, Montreal, Quebec, Canada H3G 1M8}

\author{K. G\'{o}rska}
\email{katarzyna.gorska@ifj.edu.pl}

\affiliation{H. Niewodnicza\'{n}ski Institute of Nuclear Physics, Polish Academy of Sciences, Division of Theoretical Physics, ul. Eliasza-Radzikowskiego 152, PL 31-342 Krak\'{o}w, Poland}

\author{A. Horzela}
\email{andrzej.horzela@ifj.edu.pl}

\affiliation{H. Niewodnicza\'{n}ski Institute of Nuclear Physics, Polish Academy of Sciences, Division of Theoretical Physics, ul. Eliasza-Radzikowskiego 152, PL 31-342 Krak\'{o}w, Poland}

\author{F. H. Szafraniec}
\email{fhszafra@im.uj.edu.pl}

\affiliation{Instytut Matematyki, Uniwersytet Jagiello\'{n}ski, ul. \L ojasiewicza 6, PL 30 348 Krak\'ow, Poland
}



\begin{abstract}
Following the lines of the recent paper of J.-P. Gazeau and F. H. Szafraniec [J. Phys. A: Math. Theor. {\bf 44}, 495201 (2011)], we construct here three types of coherent states, related to the Hermite polynomials in a complex variable which are orthogonal with respect to a non-rotationally invariant measure.  We investigate relations between these  coherent states and  obtain the relationship between them  and the squeezed states of quantum optics. We also obtain a second realization of the canonical coherent states in the Bargmann space of analytic functions, in terms of a squeezed basis. All this is done in the flavor of the classical approach of V. Bargmann [Commun. Pur. Appl. Math. {\bf 14}, 187  (1961)].

\end{abstract}

\maketitle


\section{Introduction}

\subsection{Motivation}

The subject of coherent states has now been developing for many decades, both in physics and mathematics. Consequently, the field has  a rich history and is still an active area of current research. The recent monographs  \cite{bSTAli00} and \cite{bJPGazeau09}, a well as the older treatise \cite{bperel}, are exhaustive sources of information on the subject, containing extensive references to the literature.

While in  the physical literature the origins of the subject go back to the 1926 paper of Schr\"odinger \cite{schroed},  what situates our present investigation within coherent states  is V. Bargmann's seminal paper \cite{VBargmann61} where coherent states somehow appeared implicitly. Here we show that even the squeezed states of quantum optics (see, for example, \cite{KWodkiewicz76,yuen}) fit in with the Bargmann construction. In fact, coherent and squeezed states may be obtained by parallel constructions following the same postulates. Direct inspiration of the present work is twofold: the well known general construction of coherent states (see, for example, \cite{bJPGazeau09}) and the approach of \cite{AHorzela12, AHorzela13}, the latter  emphasizing  the role played by the Segal-Bargmann transform in realizing  Klauder's postulates for coherent states. Here we go much further and investigate both the details of the construction as well as the mathematical and physical properties of a particular family of generalized coherent states  defined as a set of vectors in a Hilbert space $\mathcal H$,
\begin{equation}
\eta_{z} = \sum_{n=0}^{\infty}\varPhi_{n}(z)e_{n},
\label{eq1.1}
\end{equation}
the  $\varPhi_n(z)$ being an orthonormal set of functions, with respect to a given measure,  $z\in D \subset \mathbb C$ and the $e_n$ constituting an orthonormal basis of the Hilbert space.

The motivation of our paper comes from page 7 of \cite{JPGazeau11} where the measure used to obtain a resolution of the identity for  the coherent states \cite{bJPGazeau09} is not rotationally invariant, though the polynomials involved are still the Hermite polynomials. While in \cite{JPGazeau11} the emphasis is rather on locating the so obtained states in a non-commutative geometrical setting, in the present paper we pay more attention to their quantum optical relevance.

Within the general context of the definition of coherent states in (\ref{eq1.1}), we shall also, more specifically, look at what are known in the quantum optical literature as {\em non-linear coherent states}
(see, e.g., \cite{JPGazeau99,siv}) which are defined as
\begin{equation}\label{eq1.2}
\eta_{z} = {\mathcal N}^{-1/2}(|z|^2)\sum\limits_{n=0}^{\infty}{\displaystyle\frac{z^{n}}{\sqrt{\rho_{n}}}}e_{n}, \quad z\in D\; .
\end{equation}
Here $\{\rho_n\}_{n=0}^{\infty}$ is a sequence of positive numbers such that
\begin{enumerate}
\item[(i)] the series $\sum_{n=0}^{\infty}|z|^{2n}/\rho_{n}={\mathcal N}(|z|^2)$ has a non-zero radius of convergence $R$; then $D$ is a disc of radius $R$ centered at $0$,  implying that the coherent states $\eta_{z}$ are normalizable.
\end{enumerate}
 Since $\{e_{n}\}_{n=0}^{\infty}$ is an orthonormal basis of the Hilbert space ${\mathcal H}$, normalizability ensures that the the definition \eqref{eq1.2} makes sense. In addition we require that the  family of vectors $\eta_{z}$, $z\in D$, satisfy the following conditions:
\begin{enumerate}
\item[(ii)] the $\eta_{z}$ are continuous in the label $z$, \textit{i.e.}, $z^{\prime}\rightarrow z$ implies $\eta_{z^{\prime}} \rightarrow \eta_{z}$;
\item[(iii)] there exists $W(|z|^2)>0$, $z\in D$, such that the so-called  resolution of the identity condition
\begin{equation}
\frac{1}{\pi} \int_{D} (\eta_{z}\otimes \eta_{z})\, W(|z|^2) {\rm d}z = \mathbf{1}
\label{eq1.3}
\end{equation}
holds in the weak operator sense; here $(\eta_{z}\otimes \eta_{z}) x \okr \langle c_{z}, x\rangle \eta_{z}$ and $\langle\, \boldsymbol{\cdot}, \boldsymbol{-}\rangle \,$ stands for the inner product in $\mathcal{H}$ (for which we adopt the mathematicians' convention linearity in the first and antilinearity in the second term).
\end{enumerate}

We ought to emphasize here that while some of the coherent states we are dealing with are may not be new, the novel aspect of our treatment is in the mathematical representations of these states set up on  Hermite polynomials and related holomorphic functions, the previous ones to be considered as functions of a complex variable as well. While such polynomials have been observed to arise in the physical literature before,  in connection with squeezed states (see, for instance \cite{gongarv,kral,yuen}), we investigate here their deeper mathematical properties. Of particular interest is the representation  of the standard or canonical coherent states in a {\em squeezed basis} (see (\ref{eq5.10})) and of the squeezed coherent states in the Fock-Bargmann basis obtained in this manner, see (\ref{eq5.11}).

\subsection{Some words about the content of the paper}

In Section 2, we give a short review of Hermite polynomials in a complex variable. We present the holomorphic space, equipped with a non-rotational measure, on which they are orthogonal. In Section 3, we introduce a Hilbert space of entire functions and next we find the unitary mapping between this Hilbert space and the holomorphic space exhibited in Section 1. In Section 4 we construct three types of generalized coherent states. A construction of  squeezed states, within the present formalism, is presented in Section 5.  In Section 6, we consider the ladder operators and a new type of a Rodrigues formula for Hermite polynomials. We conclude in Section 7 with some comments and point out some possible future directions of work.


\section{Hermite polynomials in a complex variable}\label{sec-herm-polyn}

The Hermite polynomials
\begin{equation}\label{eq2.1}
 H_n(z) = n! \sum_{m=0}^{[n/2]} \frac {(-1)^m (2z)^{n-2m}}{m!\; (n-2m)!}, \quad z = x + {\rm i} y,
\end{equation}
considered as functions of the complex variable are orthogonal \cite{SJLvanEijndhoven90} in the space
\begin{equation}\label{eq2.2}
\mathcal{L}^{2}\big(\mathbb{C},\, \mathrm{e}^{-(1-s)x^2 - (\frac 1s -1)y^2}\, \mathrm{d}x\, \mathrm{d}y\big)
\end{equation}
where  $0 < s < 1$ is a parameter. Denoting their closure within the space \eqref{eq2.2}  by $\mathcal{H}_{\mathrm{hol}}^{s}$, it is seen to consist  of all the holomorphic functions in this space, equipped  with the inner product inherited from the space \eqref{eq2.2}.  With
\begin{equation*}
b_{n}(s) \okr \frac{\pi \sqrt{s}}{1-s}\left( 2\, \frac {1+s}{1-s}\right)^n n!\, ,
\end{equation*}
the orthogonality reads explicitly as
\begin{equation}\label{eq2.3}
\langle H_{m}, H_{n}\rangle_{\mathrm{hol}} = \int_{\mathbb C} \overline{H_m (z)} H_n (z)\; \mathrm{e}^{-(1-s)x^2 - (\frac 1s -1)y^2}\, \mathrm{d}x \mathrm{d}y = b_{n}(s)\,\delta_{mn}.
\end{equation}
After redefining the polynomials as
\begin{equation}\label{eq2.4}
H_{s, n} \okr  \frac{ H_{n}}{\sqrt{b_n(s)}}, \quad n =0, 1, 2, \ldots \; ,
\end{equation}
one has
\begin{equation*}
\langle H_{s, m}, H_{s, n} \rangle_{\mathrm{hol}} = \delta_{m n}
 \end{equation*}
which makes the sequence $\{H_{s, n}\}_{n}$ orthonormal. Furthermore, this set of vectors turns out to be complete, hence an orthonormal basis in $\mathcal{H}_{\mathrm{hol}}^{s}$.

The space $\mathcal{H}_{\mathrm{hol}}^{s}$ is a reproducing kernel Hilbert space with the kernel
\begin{align*}\label{eq5}
\begin{split}
K_{s}(z, w) &= \sum_{n=0}^\infty H_{s, n}(z)\, \overline{H_{s, m}(w)} = \frac {1-s}{\pi\sqrt{s}} \sum_{n=0}^\infty \left(\frac {1-s}{1+s}\right)^n \frac {H_n (z)\, \overline{H_n (w)}}{2^n n!}\nonumber\\[0.7\baselineskip]
&= \frac {1-s^2}{2\pi s} \exp \left[ -\frac {(1-s)^2}{4s} (z^2 + \bar{w}^2) + \frac {1-s^2}{2s}z \bar{w}\right],\quad z,w\in\mathbb C
\end{split}
\end{align*}
and the inner product defined by this kernel coincides with $\langle\,\cdot\,,\,-\,\rangle_{\rm hol}$.
The space $\mathcal{H}_{\mathrm{hol}}^{s}$ was introduced in \cite{SJLvanEijndhoven90}, see also  \cite{FHSzafraniec98}.


\section{A transformation of the Segal-Bargmann  type}\label{sec-seg-barg}

\subsection{Some basic facts about the Bargmann space and the canonical coherent states}

By the classical Bargmann space $\mathcal{H}_{\rm{hol}}$ we mean those functions in $\mathcal{L}^{2}(\mathbb C, \pi^{-1}{\mathrm{e}^{-\vert z\vert^2}}\mathrm{d}x\,\mathrm{d}y )$ which are holomorphic, or equivalently, the closure of all polynomials $\mathbb{C}[Z]$ in $\mathcal{L}^{2}(\mathbb C, \pi^{-1}{\mathrm{e}^{-\vert z\vert^2}}\mathrm{d}x\,\mathrm{d}y )$. The sequence
\begin{equation}\label{eq3.1}
\varPhi_{n} (z) \okr \frac {z^n}{\sqrt{n!}}\;,\quad z\in\mathbb C, \quad n =0,1,2, \ldots,
\end{equation}
of monomials is an orthonormal basis in $\mathcal{H}_{\rm{hol}}$. Using the physicists' convention we shall call this the {\em Fock-Bargmann basis\/.}

The associated reproducing kernel for $\mathcal{H}_{\mathrm{hol}}$ is of the form
\begin{equation*}
K(z, \overline{w}) = \sum_{n = 0}^{\infty} \varPhi_{n}(z) \overline{\varPhi_{n}(w)} = {\rm e}^{z \overline{w}},\quad z,w\in\mathbb C.
\end{equation*}
The {\em normalized} CS (or in other words, the {\em canonical coherent states}) are customarily defined as
\begin{equation}\label{eq3.2}
\eta_{\bar{z}} = \frac1{\sqrt{K(z, \bar{z})}}\, \sum_{n=0}^{\infty} \overline{\varPhi_{n}(z)}\, \varPhi_{n} = {\rm e}^{-\frac {\vert z\vert^2}2} \sum_{n=0}^{\infty} \frac{\overline{z}^n}{\sqrt{n!}}\varPhi_n,\quad z\in\mathbb C\; .
\end{equation}
with resolution of the identity
\begin{equation*}
\int_{\mathbb{C}} (\eta_{\bar{z}}\otimes \eta_{\bar{z}})\; \frac {\mathrm{d}x\,\mathrm{d}y}{\pi} = I_{\mathcal{H}_{\mathrm{hol}}}, \quad z=x+{\rm i}y
\end{equation*}
understood in the week operator sense.
\begin{remark}\label{6Aug13-3}
Let us point out here that in the physical literature it is customary to write the canonical coherent states, in the Dirac notation, as
$$
\vert\alpha\rangle =   {\rm e}^{-\frac {\vert \alpha\vert^2}2} \sum_{n=0}^{\infty} \frac{\alpha^n}{\sqrt{n!}}\vert n\rangle,\quad \alpha\in\mathbb C\; .$$
Here we have chosen to represent the  vectors $\vert n\rangle$ (of the so-called number basis) by the elements of the  Fock-Bargmann  basis.
\end{remark}

\subsection{The transforms}

Introducing the Hilbert space $\mathcal{X}^{s}_{\rm{hol}}$ of entire functions $f$ such that
\begin{equation*}
\int_{\mathbb{C}} |f(z)|^2 {\rm e}^{s x^2 - \frac{1}{s} y^2} {\rm d}x {\rm d} y < \infty, \quad z = x + {\rm i}y
\end{equation*}
and in view of  \eqref{eq2.3}, we can define an orthonormal basis in $\mathcal{X}^{s}_{\rm{hol}}$ in the form
\begin{equation}\label{eq3.3}
h_{s, n}(z) \okr {\rm e}^{-\frac{z^2}{2}} H_{s, n}(z),\quad z\in\mathbb C\; ,
\end{equation}
with $H_{s, n}$ as in \eqref{eq2.4}. Analogously, we can introduce the Hilbert space $\bar{\mathcal{X}}^{s}_{\rm{ahol}} \okr \overline{\mathcal{X}^{s}_{\rm{hol}}}$ of antiholomorphic functions with the orthonormal basis $\{\bar{h}_{s, n}\}_{n=0}^{\infty}$, $\bar{h}_{s, n}(z) \okr \overline{h_{s}(z)}$, $z\in\mathbb{C}$.

\subsubsection{From $\mathcal{X}^{s}_{\rm{hol}}$ to $\mathcal{H}_{\rm{hol}}$}

Let us now find a unitary mapping of $\mathcal{X}^{s}_{\rm{hol}}$ onto $\mathcal{H}_{\rm{hol}}$. We can write the well-known formula of the generating function for the Hermite polynomials in the following form
\begin{equation*}
\sum_{n=0}^{\infty} \frac{\left(\sqrt{\frac{\epsilon}{2}} z\right)^n}{n!} H_{n}(w) = {\rm e}^{\sqrt{2 \epsilon} z w - \frac{\epsilon}{2} z^2},\quad z,w\in\mathbb C.
\end{equation*}
Setting $\epsilon = \frac{1-s}{1+s}$, using \eqref{eq3.3} and \eqref{eq2.4}, and expressing $H_{n}(w)$ in terms of $h_{s, n}(w)$ we get
\begin{equation}\label{eq3.4}
\sum_{n=0}^{\infty} \varPhi_{n}(z) \overline{h_{s, n}(w)} = \left(\frac{1-s}{\pi \sqrt{s}}\right)^{\frac{1}{2}} \exp\big[\sqrt{2 \epsilon} z \bar{w} - \dfrac{1}{2} (\epsilon z^2 + \bar{w}^2)\big]
\end{equation}
and in view of the orthogonality relation \eqref{eq2.4}, we get the transformation $\varPhi_{n} = A h_{s, n}$,
\begin{equation}\label{eq3.5}
\varPhi_{n}(z) = (A h_{s, n}) (z) = \int_{\mathbb{C}} A(z, \bar{w}) h_{s, n}(w) {\rm e}^{s u^2 - \frac{1}{s} v^2} {\rm d}u {\rm d} v, \quad w = u + {\rm i} v
\end{equation}
with the kernel
\begin{equation}\label{eq3.6}
A(z, \bar{w}) = \left(\frac{1-s}{\pi \sqrt{s}}\right)^{\frac{1}{2}} {\rm e}^{\sqrt{2 \epsilon} z \bar{w} - \frac{1}{2} (\epsilon z^2 + \bar{w}^2)},\quad z,w\in\mathbb C.
\end{equation}
One readily verifies that
\begin{equation}\label{eq3.7}
\int_{\mathbb{C}} A(z, \overline{q}) \overline{A(w, \overline{q})} {\rm e}^{s u^2 - \frac{1}{s} v^2} {\rm d}u {\rm d}v = {\rm e}^{z \bar{w}}, \quad q = u + {\rm i} v.
\end{equation}

\subsubsection{Unitarity of A}

The operator $A$ is unitary. This can be proved by showing that range of $A$ is dense in $\mathcal{H}_{\rm{hol}}$. Following  \cite{VBargmann61}, let us take the function $f_{\overline{w}}\in\mathcal{X}^{s}_{\rm hol}$, of the form $f_{\overline{w}}(q) = \overline{A(w, \overline{q})}$, $w \in \mathbb{C}$. Then, \eqref{eq3.5} and \eqref{eq3.7} give us
\begin{equation*}
\int_{\mathbb{C}} A(z, \overline{q}) f_{\overline{w}}(q) {\rm e}^{s u^2 - \frac{1}{s} v^2} {\rm d}u {\rm d}v = e^{z \overline{w}} = K_{\overline{w}}(z), \quad q=u + {\rm i}v.
\end{equation*}
As shown in \cite{AHorzela12, VBargmann61} the functions $K_{\overline{w}}$ are complete in $\mathcal{H}_{\rm hol}$, this concludes the proof of unitarity of $A$.

\subsubsection{From $\mathcal{H}_{\rm{hol}}$ to $\mathcal{X}^{s}_{\rm{hol}}$}

The unitarity of the operator $A$ implies existence of the inverse operator $A^{-1}$ which maps $\mathcal{H}_{\rm{hol}}$ onto $\mathcal{X}^{s}_{\rm{hol}}$. Moreover, Eqs.~\eqref{eq3.4}, \eqref{eq3.5}, \eqref{eq3.6} suggest the relation $A^{-1} f = W f$, more explicitly
\begin{equation}\label{eq3.8}
(W f)(w) = \int_{\mathbb{C}} \overline{A(z,\overline{w})}\; f(z)\; {\rm e}^{-\frac{|z|^2}{2}} \frac{{\rm d}x {\rm d} y}{\pi}, \quad z = x + {\rm i} y.
\end{equation}
Let us now prove the relation $A^{-1}f = W f$. We start with a function $g = A (W f)\in \mathcal{H}_{\rm hol}$. Indeed, employing \eqref{eq3.5}, \eqref{eq3.7} and \eqref{eq3.8} we get
\begin{align*}
\begin{split}
g(w) &= \int_{\mathbb{C}} A(w, \overline{q}) \left[\int_{\mathbb{C}} \overline{A(z, \overline{q})} f(z) {\rm e}^{-\frac{\vert z\vert^2}{2}} \frac{{\rm d}x {\rm d}y}{\pi} \right] {\rm e}^{\; su^2 - \frac{1}{s} v^2}  {\rm d}u {\rm d}v \nonumber\\[0.4\baselineskip]
&= \int_{\mathbb{C}} f(z) \left[\int_{\mathbb{C}} A(w, q) \overline{A(z, \overline{q})} {\rm e}^{\; su^2 - \frac{1}{s} v^2}  {\rm d}u {\rm d}v\right] {\rm e}^{-\frac{\vert z\vert^2}{2}} \frac{{\rm d}x {\rm d}y}{\pi} \nonumber\\[0.4\baselineskip]
&= \int_{\mathbb{C}} K(w, \overline{z}) f(z) {\rm e}^{-\frac{\vert z\vert^2}{2}} \frac{{\rm d}x {\rm d}y}{\pi} = f(w), \quad z = x + {\rm i} y, \quad q = u + {\rm i} v.
\end{split}
\end{align*}
The integral kernel corresponding to the transformation $A^{-1}$ is given by
\begin{equation*}
A^{-1}(z, \overline{w}) = \overline{A(z, \overline{w})} = \left(\frac{1-s}{\pi\sqrt{s}}\right)^{\frac{1}{2}} {\rm e}^{\; \sqrt{2 \epsilon} \overline{z} w - \frac{1}{2}(\epsilon \overline{z}^2 + w^2)}.
\end{equation*}


\section{Coherent states}

\begin{remark}\label{rem1}
Let us recall the construction of a  reproducing kernel Hilbert space through its prospective basis. Consider a sequence $\{\varphi_{n}\}_{n=0}^{\infty}$ of complex functions on $X$ which satisfy the following two conditions \cite{AHorzela12}:
\begin{itemize}
\item[(i)] $\sum_{n} |\varphi_{n}(x)|^{2} < \infty$, $x\in X$
\item[(ii)] $\{\alpha_{n}\}_{n=0}^{\infty} \in l_{\infty}^{2}$ and $\sum_{n=0}^{\infty} \alpha_{n} \varphi_{n}(x) = 0$ for all $x$ imply that all $\alpha_{n}$'s are zeros.
\end{itemize}
Then the kernel $K: X\times X \to \mathbb{C}$ related to $\varphi_{n}$ given as
\begin{equation}\label{eq4.1}
K(x, y) \okr \sum_{n=0}^{\infty} \varphi_{n}(x) \overline{\varphi_{n}(y)}
\end{equation}
is positive definite and  $\{\varphi_{n}\}_{n=0}^{\infty}$ constitutes a basis of its reproducing kernel Hilbert space. Conversely,  if $\{\varphi_{n}\}$ is an orthonormal basis in the reproducing kernel Hilbert space determined by $K$, then the  formula \eqref{eq4.1} defining $K$ turns into equality and says that $K$ decomposes according to \eqref{eq4.1}, where the series converges uniformly on any subset of $X$ on which the function $K(\,\cdot\,,\,\cdot\,)$ is bounded.
\end{remark}

Although constructing the coherent states we have insisted on having a resolution of the identity \eqref{eq1.3} to hold, which is necessary for various physical applications, it is possible to define coherent states using the properties of the reproducing kernel alone. An approach to the construction of coherent states not requiring the presence of any $\mathcal{L}^{2}$ space, yet \underbar{with} an intrinsic resolution of the identity built into it, has been successfully accomplished in \cite{AHorzela12} and slightly improved in \cite{AHorzela13}. The resolution of the identity is usual a consequence of the way of coherent states are introduced; for instance in \cite{bSTAli00} it stems from addition assumptions (cf. Chapter 6 of \cite{bSTAli00}) attached to the construction which is assigned to the direct integral Hilbert spaces formalism (cf. Chapter 5 of \cite{bSTAli00}) while in Perelomov's book \cite{bperel} it comes out as a result of group theoretical approach. There is a vast literature concerning  different possible constructions of CS, for which an updated and extensive list of references can be found in the forthcoming second edition of \cite{bSTAli00}; in addition to that a recent paper \cite{odzhor} concerning the matter, in a geometrical setting,  deserves to be mentioned here. 

According to the philosophy put forward in \cite{AHorzela12,AHorzela13}, coherent states built therein require exclusively two sets of parameters:
\begin{itemize}
\item[(a)] a sequence $\{\varphi_{n}\}_{n}$ of functions of $x$ satisfying conditions (i) and (ii) above,
\item[(b)] a sequence $\{e_{n}\}_{n}$ constituting an orthonormal basis of $\mathcal{H}$.
\end{itemize}
Then the coherent states, determined by the parameters (a), (b) and corresponding to the kernel \eqref{eq4.1}), are defined as
\begin{equation*}
\eta_{\bar{x}} = N(x) \sum_{n=0}^{\infty} \overline{\varphi_{n}(x)}\, e_{n}, \quad N(x) = K^{-1/2}(x, x), \quad x\in X.
\end{equation*}
Since physical coherent states are state vectors of quantum systems, they have to be normalized and hence the introduction of the normalization factor $N(x)$ above. However, mathematically this is not necessary and not having it does not influence the existence of a resolution of the identity; moreover, if the functions $\varphi (x)$ happen to be holomorphic, this property is then inherited by the non-normalized coherent states (see footnote {\sf b} in \cite{AHorzela13}).

Specifying the parameters in (a) and (b) above, using the polynomials and functions defined in Sections \ref{sec-herm-polyn} and \ref{sec-seg-barg}, we get nine possible types of coherent states, all of them more or less natural, as listed in Table~\ref{tab1} below. (Observe, however, that in all these cases a resolution of the identity is satisfied).
\begin{table}[!h]
\begin{center}
\begin{tabular}{c | c c c c}
\; & $\quad\varphi_{n}$ & $\quad e_{n}$ & $\quad N$ & $\quad C_{\bar{z}}$ \\[0.3\baselineskip] \hline\hline
1. & $\quad\varPhi_{n}(z)$ & $\quad\varPhi_{n}$ & $\quad {\rm e}^{-|z|^{2}/2}$  & $\quad {\rm e}^{-|z|^{2}/2} \sum_{n=0}^{\infty} \overline{\varPhi_{n}(z)} \varPhi_{n}$ \\[0.5\baselineskip]
2. & $\quad\varPhi_{n}(z)$ & $\quad H_{s, n}$ & $\quad {\rm e}^{-|z|^{2}/2}$ & $\quad {\rm e}^{-|z|^{2}/2} \sum_{n=0}^{\infty} \overline{\varPhi_{n}(z)} H_{s, n}$ \\[0.5\baselineskip]
3. & $\quad\varPhi_{n}(z)$ & $\quad h_{s, n}$ & $\quad {\rm e}^{-|z|^{2}/2}$ &  $\quad {\rm e}^{-|z|^{2}/2} \sum_{n=0}^{\infty} \overline{\varPhi_{n}(z)} h_{s, n}$ \\[0.3\baselineskip] \hline\hline
4. & $\quad H_{s, n}(z)$ & $\quad\varPhi_{n}$ & $\quad \kappa_{s; z}(z)$ & $\quad \big(\ulamek{2\pi s}{1-s^{2}}\big)^{\!\frac{1}{2}} \exp\left[- \frac{1-s^{2}}{4 s} |z|^{2} + \frac{(1-s)^{2}}{8 s}(z^{2} + \bar{z}^{2})\right] \sum_{n=0}^{\infty} \overline{H_{s, n}(z)} \varPhi_{n}$ \\[0.5\baselineskip]
5. & $\quad H_{s, n}(z)$ & $\quad H_{s, n}$ & $\quad \kappa_{s; z}(z)$ & $\quad \big(\ulamek{2\pi s}{1-s^{2}}\big)^{\!\frac{1}{2}} \exp\left[- \frac{1-s^{2}}{4 s} |z|^{2} + \frac{(1-s)^{2}}{8 s}(z^{2} + \bar{z}^{2})\right] \sum_{n=0}^{\infty} \overline{H_{s, n}(z)} H_{s, n}$ \\[0.5\baselineskip]
6. & $\quad H_{s, n}(z)$ & $\quad h_{s, n}$ &$\quad \kappa_{s; z}(z)$ & $\quad \big(\ulamek{2\pi s}{1-s^{2}}\big)^{\!\frac{1}{2}} \exp\left[- \frac{1-s^{2}}{4 s} |z|^{2} + \frac{(1-s)^{2}}{8 s}(z^{2} + \bar{z}^{2})\right] \sum_{n=0}^{\infty} \overline{H_{s, n}(z)} h_{s, n}$
\\[0.3\baselineskip] \hline\hline
7. & $\quad h_{s, n}(z)$ & $\quad\varPhi_{n}$ & $\quad \tilde{\kappa}_{s; z}(z)$ & $\quad \big(\ulamek{2\pi s}{1-s^{2}}\big)^{\!\frac{1}{2}} \exp\left[- \frac{1-s^{2}}{4 s} |z|^{2} + \frac{1+s^{2}}{8 s}(z^{2} + \bar{z}^{2})\right] \sum_{n=0}^{\infty} \overline{h_{s, n}(z)} \varPhi_{n}$ \\[0.5\baselineskip]
8. & $\quad h_{s, n}(z)$ & $\quad H_{s, n}$ & $\quad \tilde{\kappa}_{s; z}(z)$ & $\quad \big(\ulamek{2\pi s}{1-s^{2}}\big)^{\!\frac{1}{2}} \exp\left[- \frac{1-s^{2}}{4 s} |z|^{2} + \frac{1+s^{2}}{8 s}(z^{2} + \bar{z}^{2})\right] \sum_{n=0}^{\infty} \overline{h_{s, n}(z)} H_{s, n}$ \\[0.5\baselineskip]
9. & $\quad h_{s, n}(z)$ & $\quad h_{s, n}$ &$\quad \tilde{\kappa}_{s; z}(z)$ & $\quad \big(\ulamek{2\pi s}{1-s^{2}}\big)^{\!\frac{1}{2}} \exp\left[- \frac{1-s^{2}}{4 s} |z|^{2} + \frac{1+s^{2}}{8 s}(z^{2} + \bar{z}^{2})\right] \sum_{n=0}^{\infty} \overline{h_{s, n}(z)} h_{s, n}$
\end{tabular}
\end{center}
\caption{{\small In the above} \\
$\kappa_{s; z}(z) \okr \sqrt{\ulamek{2\pi s}{1-s^{2}}} \exp\left[- \frac{1-s^{2}}{4 s} |z|^{2} + \frac{(1-s)^{2}}{8 s}(z^{2} + \bar{z}^{2})\right]$, $\tilde{\kappa}_{s; z}(z) \okr  \sqrt{\ulamek{2\pi s}{1-s^{2}}} \exp\left[- \frac{1-s^{2}}{4 s} |z|^{2} + \frac{1+s^{2}}{8 s}(z^{2} + \bar{z}^{2})\right]$.
}
\label{tab1}
\end{table}

The first group of three coherent states above are just the canonical coherent states, realized as elements in the Hilbert spaces spanned by the vectors $\varPhi_{n}, H_{s, n}$ and $h_{s, n}$, respectively, while the second and the third groups are two potentially new types of coherent states defined on these same spaces.

The ladder operators $b^{-}$ and $b^{+}$ satisfying the commutation relation $[b^{-}, b^{+}] = 1$, in the abstract setup act on the orthonormal basis $e_{n}$ in the following way
\begin{equation}\label{eq4.2}
b^{-} e_{n} = \sqrt{n}\; e_{n-1}, \quad b^{+} e_{n} = \sqrt{n+1}\; e_{n+1}.
\end{equation}
Specifying the Hilbert space and the basis we distinguish between  three  cases. For the basis of monomials in $\mathcal{H}_{\rm hol}$ these operators are represented as derivation and multiplication operators, respectively \cite{VBargmann61} and the canonical coherent states $\eta_{\overline{z}}$ (see no. 1. in Tab. \ref{tab1}) are the eigenstates of the lowering operator $b^{-}$ corresponding to the eigenvalue $\overline{z}$. The explicit forms of the ladder operators in $\mathcal{H}^{s}_{\rm hol}$ which act in accordance with \eqref{eq4.2} can be calculated from the recursion formulae for the  Hermite polynomials $2n H_{n-1} = H^{\;\prime}_{n}$ and $H_{n+1} = 2 z H_{n} - H_{n}^{\,\prime}$:
\begin{equation*}
A^{-}_{s} = \left[\frac{1+s}{2(1-s)}\right]^{\ulamek{1}{2}} \partial_{z}, \quad A^{+}_{s} = \left[\frac{1-s}{2(1+s)}\right]^{\ulamek{1}{2}} (2 z - \partial_{z}).
\end{equation*}
Similarly, on  $\mathcal{X}^{s}_{\rm hol}$ they behave like
\begin{equation*}
a^{-}_{s} = \left[\frac{1+s}{2(1-s)}\right]^{\ulamek{1}{2}} (z + \partial_{z}), \quad a^{+}_{s} = \left[\frac{1-s}{2(1+s)}\right]^{\ulamek{1}{2}} (z - \partial_{z}),
\end{equation*}
which is reminiscent of those on $\mathcal{L}^{2}(\mathbb{R})$.

Simple calculation ensure us that the coherent states in nos. 2. and 3. in Tab.~\ref{tab1} are  eigenfunctions of $A^{-}_{s}$ and $a^{-}_{s}$, respectively, corresponding to the eigenvalue $\overline{z}$ as well.


\section{The squeeze operator and squeezed coherent states}

We show how the complex Hermite polynomials $H_n (z)$ are related to the {\em squeezed coherent states} of quantum optics (see also \cite{yuen}). With $\eta_{\overline{z}}$ as in (\ref{eq3.2}), we define the {\em squeezed states} $\eta^\xi_{\overline{z}}$ is as follows
\begin{equation}\label{eq5.1}
\eta_{\bar{z}}^\xi \okr S(\xi) \eta_{\overline{z}},  \qquad S(\xi) \okr \mathrm{e}^{\xi K_+ - \overline{\xi} K_-},\quad \xi \in \mathbb C,
\end{equation}
where $K_+ , K_-$ are two of the usual generators of the $SU(1,1)$ group. These generators together with the third, $K_0$, satisfy the well-known commutation relations
\begin{equation*}
[K_- , K_+ ] = 2K_0, \quad [K_0 , K_\pm ] = \pm K_\pm.
\end{equation*}
In the Bargmann-like representation (see \cite[formula (I.3.42), p. 17]{GDattoli97}) they have the forms
\begin{equation}\label{eq5.2}
K_{+} = \ulamek{1}{2} z^{2} , \quad K_{-} = \ulamek{1}{2} \partial_{z}^{2} , \quad K_{0} = \ulamek{1}{2} \Big[z\partial_{z} + \ulamek{1}{2} \Big].
\end{equation}
The operator $S(\xi)$ is unitary and is known as the {\em squeeze operator\/.}

Thus, from \eqref{eq3.2}, we have
\begin{equation}\label{eq5.3}
\eta_{\bar{z}}^\xi = {\rm e}^{-\frac {\vert z\vert^2}2}\sum_{n=0}^\infty  \frac {\overline{z}^n}{\sqrt{n!}}\varPhi_n^\xi, \quad \text{where,} \quad  \varPhi_{n}^\xi \okr S(\xi) \varPhi_{n};
\end{equation}
as a representation of the squeezed coherent state in terms of the {\em squeezed basis} $\varPhi_n^\xi, \; n =0,1,2, \ldots\;$. Since the squeeze operator is unitary, the squeezed basis is also  orthonormal in the Bargmann space $\mathcal{H}_{\rm{hol}}$. It is clear that the squeezed coherent states states (\ref{eq3.2}) satisfy the same resolution of the identity as the canonical coherent states $\eta_{\overline{z}}$. We shall find below (see (\ref{eq5.11})) a second representation of the squeezed states in the Fock basis $\varPhi_n, \; n = 0,1,2, \ldots\; $.

\subsection{Analytic form of the squeezed basis vectors}

The  unitarity of the squeeze operator $S(\xi )$, defined by \eqref{eq5.1} and \eqref{eq5.2} implies,
\begin{equation}\label{eq5.4}
S(\xi)^{*} = S(\xi)^{-1} = S(-\xi), \qquad \xi\in\mathbb{C},
\end{equation}
 and since the vectors $\{\varPhi_n^\xi\}_{n= 0}^\infty$ also form an orthonormal basis of $\mathcal{H}_{\rm{hol}}$,  Remark \ref{rem1} allows us to write
\begin{equation}\label{eq5.5}
K(z, \overline{w}) = \sum_{n=0}^{\infty} \varPhi_{n}(z) \overline{\varPhi_{n}(w)} = \sum_{n=0}^{\infty} \varPhi^{\xi}_{n}(z) \overline{\varPhi^{\xi}_{n}(w)}.
\end{equation}

The squeezed basis vectors  $\varPhi^{\xi}_n$ coincide with the complex Hermite polynomials $H_{s, n}(z)$ times am exponential factor, as we now demonstrate.  Note that, setting $\zeta = \ulamek{\tanh (\vert\xi\vert)}{\vert \xi\vert}\xi$, the well-known {\em disentanglement formula} takes the form
\begin{equation}  \label{eq5.6}
{\rm e}^{\xi K_{+} - \overline{\xi}K_{-}} = {\rm e}^{\zeta K_{+}}\; {\rm e}^{[\log (1 -\vert\zeta\vert^2)] K_{0}}\; {\rm e}^{-\overline{\zeta} K_{-}} = {\rm e}^{\frac{\zeta}{2} z^{2}}\; {\rm e}^{\log (1 -\vert\zeta\vert^{2}) \frac{1}{2}(z\partial_{z} + \ulamek{1}{2})}\; {\rm e}^{-\frac{\bar{\zeta}}{2} \partial_{z}^{2}},
\end{equation}
which makes sense because $|\zeta| < 1$.  Employing  \eqref{eq5.2} we get
\begin{equation*}
{\rm e}^{-\overline{\zeta}K_{-}}\varPhi_{n}(z) = {\rm e}^{-\frac{\bar{\zeta}}{2} \partial_{z}^{2}}\left(\frac{z^n}{\sqrt{n!}} \right) = \sqrt{n!} \sum_{m=0}^{[n/2]}
\frac{\big(-\ulamek{\bar{\zeta}}{2}\big)^{m} z^{n - 2m}}{m! (n - 2m)!}.
\end{equation*}
Next
 \begin{equation} \label{eq5.7}
\exp\left[\log (1 - \vert\zeta\vert^{2}) \ulamek{1}{2} \big(z\partial_{z} + \ulamek{1}{2}\big)\right] z^{k} = (\sqrt{1 - \vert\zeta\vert^{2}})^{k + \frac{1}{2}} z^{k}
\end{equation}
since for any integer $k$, $z\partial_z z^k = kz^k$. Combining \eqref{eq5.6}, \eqref{eq5.7} and noting \eqref{eq2.1} we arrive at the result describing the squeezed basis analytically.
\betheo\label{theor-sq-basis}
For any non-zero $\xi \in \mathbb C$ the squeezed basis vectors are given by
\begin{equation*}
\varPhi^{\xi}_{n}(z) = (1- \vert\zeta\vert^2)^{\frac{1}{4}} {\rm e}^{\frac{\zeta}{2} z^{2}} \frac{\overline{\zeta}^{\frac{n}{2}}}{\sqrt{2^{n} n!}}\; H_{n}\Big(\sqrt{\ulamek{1- \vert\zeta\vert^2}{2\overline{\zeta}}} \, z\Big), \qquad \zeta = \ulamek{\tanh (\vert\xi\vert)}{\vert \xi\vert}\xi\; .
\end{equation*}
\entheo
Note that $0< \vert\zeta\vert < 0$.

Let us now define the {\em dilation} operator $R^{\xi}$,
\begin{equation}\label{eq5.8}
R^{\xi} f(z) = f\Big(\sqrt{\ulamek{2 \overline{\zeta}}{1-\vert\zeta\vert^{2}}}\, z\Big), \quad z \in\mathbb{C}.
\end{equation}
Writing $\zeta \okr \ulamek{1-s}{1+s}$ we get
\begin{align}\label{eq5.9}
\begin{split}
R^{\xi} \left[{\rm e}^{-|z|^{2}} \varPhi^{\xi}_{n}(z) \overline{\varPhi^{\xi}_{n}(z)}\right] &=  \kappa^{2}_{s; w}(w) H_{s, n}(w) \overline{H_{s, n}(w)} \\[0.4\baselineskip]
& =  \tilde{\kappa}^{2}_{s; w}(w) h_{s, n}(w) \overline{h_{s, n}(w)}
\end{split}
\end{align}
with $\kappa_{s; z}(z)$ and $\tilde{\kappa}_{s; z}(z)$ given in Tab.~\ref{tab1}. The formulae \eqref{eq5.9} shed further light on the relationship between the vectors $\varPhi_{n}^{\xi}$ on the one hand and  $H_{s, n}$ and $h_{s, n}$ on the other.

\subsection{Coherent states  and squeezed squeezed coherent states revisited}

In view of  \eqref{eq5.3}, the squeeze operator $S(\xi )$ can be written as
\begin{equation*}
S(\xi) = \sum_{n=0}^{\infty} \varPhi_{n}^{\xi} \otimes \varPhi_{n},
\end{equation*}
and thus on the Bargmann space $\mathcal{H}_{\rm hol}$ it becomes an integral operator with the kernel
\begin{align*}
\begin{split}
S^{\xi}(z, \bar{w}) &= \sum_{n=0}^{\infty} \varPhi^{\xi}_{n}(z)\overline{\varPhi_{n} (w)} \nonumber\\
& = \sum_{n=0}^{\infty} \frac{1}{\sqrt{n!}}(1- \vert\zeta\vert^{2})^{\frac{1}{4}} \big(\ulamek{\overline{\zeta}}{2}\big)^{\frac{n}{2}} {\rm e}^{\frac{\zeta}{2} z^2} H_{n}\Big(\sqrt{\ulamek{1- \vert\zeta\vert^2}{2\overline{\zeta}}}\, z\Big) \frac {\bar{w}^n}{\sqrt{n!}}.
\end{split}
\end{align*}
Rearranging this and using the well-known generating function for Hermite polynomials \cite[formula 8.9.57.1]{ISGradshteyn07} we get
\begin{equation*}
S^{\xi}(z, \bar{w}) = (1 - \vert\zeta\vert^2)^{\frac{1}{4}} {\rm e}^{\frac{1}{2}(\zeta z^{2} - \bar{\zeta}\bar{w}^{2})} {\rm e}^{\sqrt{1 - \vert\zeta\vert^2}\, z \bar{w}}.
 \end{equation*}
This implies, by \eqref{eq5.3}, that
\begin{equation*}
\eta_{\overline{w}}^{\xi}(z) = (1 - \vert\zeta\vert^2)^{\frac{1}{4}} {\rm e}^{- \frac{\vert w\vert^2}{2}} e^{\frac{1}{2}(\zeta z^{2} - \overline{\zeta}\bar{w}^{2})} {\rm e}^{\sqrt{1 - \vert\zeta\vert^2}\, z \bar{w}} = {\rm e}^{-\frac{\vert w\vert^{2}}{2}}S^{\xi} (z, \bar{w}).
\end{equation*}
Note also that taking the limit $s\to 1$ which is equivalent to $\zeta\to 0$ yields to the canonical Bargmann space setting as already noted in \cite[Subsection 6.3]{JPGazeau11}. Furthermore, Eq.~\eqref{eq5.3} implies
\begin{equation*}
\varPhi^{\xi}_{n}(z) = \frac{1}{\pi} \int_{\mathbb C} S^{\xi}(z, \bar{w}) \varPhi_{n}(w) {\rm e}^{-\vert w\vert^{2}} {\rm d} u {\rm d}v, \quad w = u + iv,
\end{equation*}
and since, according to \eqref{eq5.4}, the inverse operator $S(\xi )^{-1}$ has the kernel $S^{-\xi} (z, \overline{w})$,
\begin{equation*}
\varPhi_{n}(z) = \frac{1}{\pi} \int_{\mathbb C} S^{-\xi}(z, \bar{w}) \varPhi^{\xi}_{n}(w) {\rm e}^{-\vert w\vert^2} {\rm d}u {\rm d}v.
\end{equation*}

Next, we show that the canonical coherent states in $\mathcal{H}_{\rm hol}$ can be considered in two equivalent ways by rewriting \eqref{eq5.5} as
\begin{equation*}
K(\,\cdot\, , z) = \sum_{n} \overline{\varPhi_{n}(z)} \varPhi_{n} = \sum_{n} \overline{\varPhi^{\xi}_{n}(z)} \varPhi^{\xi}_{n}.
\end{equation*}
Indeed, this implies that
\begin{equation}\label{eq5.10}
\eta_{\bar{z}} = {\rm e}^{-\frac {\vert z\vert^2}2}\sum_{n=0}^{\infty} \overline{\varPhi_{n}(z)} \varPhi_{n} = {\rm e}^{-\frac {\vert z\vert^2}2}\sum_{n=0}^{\infty} \overline{\varPhi^{\xi}_{n}(z)} \varPhi^{\xi}_{n}.
\end{equation}

\begin{remark}\label{6Aug13-2}
Furthermore, we emphasize here that while both expansions (\ref{eq5.10}) represent the same canonical coherent state,  the first sum  is the standard representation in the basis $\varPhi_n$ defined by \eqref{eq3.1}. It gives the Poisson distributed probability
$$ P(n, \lambda) = e^{-\lambda}\frac {\lambda^{n}}{n!}, \qquad \lambda = \vert z \vert^2, $$
of finding an $n$-exciton state when the quantum system is in the coherent state $\eta_{\overline{z}}$. Note that $\lambda$ is also the expected number of excitons in the coherent state.  The second representation, of $\eta_{\overline{z}}$ in the squeezed basis, is new, as far as we are aware. In this representation, the  $n$-th squeezed basis state appears with the probability
$$ P(n, z, \xi) = e^{-\vert z\vert^2} \vert\varPhi^\xi_n (z )\vert^2. $$

Additionally, using this representation we can obtain a second representation for the squeezed states $\eta_{\overline{z}}^\xi$ (see (\ref{eq5.3})), in terms of the classical Bargmann space basis $\varPhi_n$. Indeed, since the $\xi \in \mathbb C$ in (\ref{eq5.10}) is arbitrary, we may also write
$$
\eta_{\bar{z}}  = {\rm e}^{-\frac {\vert z\vert^2}2}\sum_{n=0}^{\infty} \overline{\varPhi^{-\xi}_{n}(z)} \varPhi^{-\xi}_{n}.$$
Applying the squeeze operator $S(\xi)$ in (\ref{eq5.1}) to both sides of this equation and noting (\ref{eq5.4}) we obtain
\begin{equation}\label{eq5.11}
\eta_{\bar{z}}^\xi  = {\rm e}^{-\frac {\vert z\vert^2}2}\sum_{n=0}^{\infty} \overline{\varPhi^{-\xi}_{n}(z)} \varPhi_{n}.
\end{equation}
The two representations (\ref{eq5.3}) and (\ref{eq5.11}) of the squeezed coherent state $\eta_{\overline{z}}^\xi$ represent an interesting kind of duality. From the first we see that the probability of appearance of the $n$-th squeezed basis state is given by the square amplitude of the $n$-th Fock-Bargmann (or number) state, while  the second shows that the probability of appearance of the $n$-th Fock-Bargmann (or number) state is  given by the the square amplitude of the corresponding squeezed basis state (see also \cite{gongarv,kral}).
\end{remark}


\section{Ladder operators on the squeezed basis and a Rodriguez type formula}

Recall that in the Bargmann space $\mathcal{H}_{\rm hol}$ the ladder operators act as
\begin{equation*}
a^{-}\varPhi_n = \sqrt{n}\varPhi_{n-1}, \quad  a^{+}\varPhi_n = \sqrt{n+1}\varPhi_{n+1}.
\end{equation*}
On the squeezed basis they give rise to the {\em squeezed ladder operators}
\begin{equation*}
a^{-}_{\xi} \okr S(\xi)a^{-}S(-\xi) \quad \text{and} \quad a^{+}_{\xi} \okr S(\xi)a^{+} S(-\xi)\; ,
\end{equation*}
so that
\begin{equation}\label{eq6.1}
a^{-}_{\xi}\varPhi^{\xi}_{n} = \sqrt{n}\varPhi^{\xi}_{n-1}, \quad  a^{+}_{\xi}\varPhi^{\xi}_n = \sqrt{n+1}\varPhi^{\xi}_{n+1}.
\end{equation}
By standard arguments, using \eqref{eq5.6}, (see, for example, \cite{KWodkiewicz76}) we obtain
\begin{equation*}
a^{-}_{\xi} = S(\xi)a^{-}S(-\xi) = \cosh(\vert\xi\vert) a^{-} - \frac \xi{\vert\xi\vert}\sinh (\vert\xi\vert)a^{+}.
\end{equation*}
with
\begin{equation*}
\cosh(\vert\xi\vert) = \ulamek{1}{\sqrt{1 - \vert\zeta\vert^2}},\qquad \sinh(\vert\xi\vert) = \ulamek{\zeta}{\sqrt{1 - \vert\zeta\vert^2}}.
\end{equation*}
As operators on $\mathcal{H}_{\rm hol}$ they have the forms
\begin{equation*}
a^{-}_{\xi} = S(\xi)a^{-} S(-\xi) = \ulamek{1}{\sqrt{1 - \vert\zeta\vert^2}}\; (\partial_{z} - \zeta z),
\end{equation*}
and
\begin{equation*}
a^{+}_{\xi} = S(\xi)a^{+} S(-\xi) = \ulamek{1}{\sqrt{1 - \vert\zeta\vert^2}}\; (z - \bar{\zeta}\partial_{z}).
\end{equation*}

Using these operators we now obtain  an interesting Rodriguez type formula for the complex Hermite polynomials $H_n(z)$.  Indeed, from \eqref{eq6.1},
\begin{equation*}
\varPhi^{\xi}_{n} (z ) = \Big[\ulamek{(a_{\xi}^{+})^{n}}{\sqrt{n!}} \varPhi^{\xi}_{0}\Big](z) = \Big[\ulamek{(z - \bar{\zeta}\partial_{z})^{n}}{\sqrt{n! (1 - \vert\zeta\vert^2)^{n}}}\; \varPhi^{\xi}_{0}\Big](z)
\end{equation*}
with
\begin{equation}\nonumber
\varPhi^{\xi}_{0}(z) = (1 -\vert\zeta\vert^2)^{\frac{1}{4}} {\rm e}^{\frac{\zeta}{2} z^{2}}, \quad \text{and} \quad a^{-}_{\xi}\varPhi^{\xi}_{0} = 0 ,
\end{equation}
and hence, we get the formula,
\begin{equation*}
H_{n}\Big(\sqrt{\ulamek{1-\vert\zeta\vert^2}{2\overline{\zeta}}}\, z\Big) = \Big[\ulamek{2}{\bar{\zeta}(1-\vert\zeta\vert^2)}\Big]^{\frac{n}{2}} {\rm e}^{- \frac{\zeta}{2} z^{2}} (z - \bar{\zeta}\partial_{z})^{n} {\rm e}^{\frac{\zeta}{2} z^{2}}.
\end{equation*}
A simple change of variables or, equivalently, using the operator $R^{\xi}$ defined in \eqref{eq5.8}, leads to
\begin{equation*}
H_{n}(z) = {\rm e}^{-\sinh^{2}(\vert\xi\vert) z^{2}}[2\cosh^{2}(\vert\xi\vert) z - \partial_{z}]^{n} {\rm e}^{\sinh^{2}(\vert\xi\vert) z^{2}},
\end{equation*}
with the left hand side being independent of $\xi$. We get in this way a kind of a Rodriguez formula
\begin{equation*}
H_{n}(z) = {\rm e}^{-az^2} [ 2(1+a)z -\partial_{z}]^{n} {\rm e}^{a z^{2}},
\end{equation*}
which holds for both for real and complex $z$. Here $a$ can be considered as a ghost parameter which could be real or complex. For $a = -1,  -\frac 12$ or $0$ we get back three well known formulae (see \cite{GBArfken01, KOldham09}).

\section{Conclusion}

The thrust of this paper has been first to exploit the analytic properties of the Hermite polynomials $H_{n}$ in a complex variable to obtain two Hilbert spaces of analytic functions, $\mathcal{H}_{\rm{hol}}^s$ and $\mathcal{X}^{s}_{\rm{hol}}$ and to relate them to the Bargmann space of analytic functions $\mathcal{H}_{\rm{hol}}^s$. On these spaces we suggested constructions of sets of coherent states, of which one set coincides with the canonical coherent states, while the other two sets are potentially new types of coherent states. The second problem addressed here has been a study of the analytic properties of the squeezed coherent states in quantum optics, realized as vectors in the Bargmann space. It would be interesting to study further the properties of the coherent states in rows 4 - 9 of  Table 1. A second problem could be to analyze physically the second representation of the canonical coherent states in (\ref{eq5.10}).

Finally we remark that without any difficulty our investigations and the results can just as easily be carried out to the antiholomorphic setting.

\section*{Acknowledgements}

Research of the fourth author was supported by the MNiSzW grant NN201 1546438. The work of STA was partially supported by an NSERC grant


\end{document}